\documentstyle{article}

\def\be{\begin{eqnarray}}
\def\ee{\end{eqnarray}}
\def\nn{\nonumber}

\hoffset= - 1.0in         

\begin{document}

\hfill ITEP/TH-25/08

\bigskip

\centerline{\Large{
From Simplified BLG Action to the First-Quantized M-Theory
}}

\bigskip

\centerline{A.Morozov}

\bigskip

\centerline{\it ITEP, Moscow, Russia}

\bigskip

\centerline{ABSTRACT}

\bigskip

{\footnotesize
Concise summary of the recent progress in the search for
the world-volume action for multiple M2 branes.
After a recent discovery of simplified version of BLG
action, which is based on the ordinary Lie-algebra
structure, does not have coupling constants and
extra dynamical fields, attention should be switched
to the study of M2 brane dynamics.
A viable brane analogue of Polyakov formalism and
Belavin-Knizhnik theorem for strings can probably
be provided by Palatini formalism for 3d (super)gravity.
}

\bigskip

\bigskip

In the context of general string theory \cite{GST}
a variety of string and superstring models,
linked by a number of duality relations, is naturally
unified in a hypothetical "M-theory" \cite{Mth},
which in its most naive perturbative phase is represented
by "fundamental (super)membranes".
The BLG action \cite{BL,G} resolves the long-standing
puzzle \cite{JS1} of finding the 3d Lagrangian with
appropriate superconformal symmetry and thus opens a way
to developing the first-quantized theory of M2 branes.
This implies that M-theory can now be studied in the same
constructive manner as bosonic and super-strings in 1980's
\cite{Po,BK,BKoth}.
Naturally, this ground-breaking achievement attracts enormous
attention \cite{BLGfirst}-\cite{triple}, and some minor drawbacks
of original analysis in \cite{BL,G} are now fully cured.

The main obvious difficulty of original BLG formulation
was its reliance upon sophisticated 3-algebra
(quantum Nambu bracket) structure \cite{3alg}
-- new for the fundamental physical considerations.
The lack of experience and intuition about this structure
caused certain confusion at the early stages:
original BLG action was written only for an artificial(?)
example of $SO(4)$ symmetry, problems were discovered with
straightforward generalizations to other groups and even
doubts appeared about the very existence of BLG action for
the stack of $N$ M2 branes with arbitrary $N$, which would
require promotion of $SO(4)$ to $SU(N)$.
The key step in overcoming this problem was analysis of the
M2 $\rightarrow$ D2 conversion in \cite{Mu}, which linked
the 3-algebra structure to conventional Lie algebras,
governing Yang-Mills and D-brane theories.
Based on this analysis, in \cite{M} a "simplified" BLG action
was introduced, which makes use of the Lie-algebra structure
only (i.e. is based on "reducible-to-Lie-algebra" Nambu bracket
of \cite{triNa}, see eq.(\ref{Nabr}) below).
The only new ingredient, distinguishing this version of
BLG action from the ones familiar from string/brane studies
was a pair of extra color-less octuplets $\varphi^I$
and $\chi^A$.
While very simple, this suggestion had serious problems as it was,
originated from degeneracy of the underlying Nambu bracket
and the lack of total antisymmetry of 3-algebra structure constants:
this made original supersymmetry proof of \cite{BL} unapplicable
and the action in \cite{M} potentially non-supersymmetric.
Thus it was meant to be a toy-example, showing the direction
to eliminate unnecessary(?) elements of the BLG construction,
but still possessing some extra fields and requiring some further
tuning.
A natural next step was to look at a central extension,
lifting degeneracy of Nambu bracket \cite{Gu}
-- and this was finally done in a triple of wonderful papers
\cite{triple}.
They resolved the discrepancy between \cite{M} and BLG approach
in an  elegant way, by changing the nature of the extra
fields $\varphi,\chi$: they are actually auxiliary, non-dynamical
variables.
Kinetic terms
$(\partial_\mu\varphi^I)^2$ and $\bar\chi^A\hat\partial\chi^A$
of \cite{M} are substituted in \cite{triple} by
$\partial_\mu\tilde\varphi^I\partial_\mu\varphi^I$ and
$\frac{1}{2}\bar{\tilde\chi}^A\hat\partial\chi^A
+ \frac{1}{2}\bar{\chi}^A\hat\partial\tilde\chi^A$
where $\tilde\varphi,\tilde\chi$ is still another
pair of color-less octuplets which do not appear anywhere else
in the action and thus serve as Lagrange multipliers,
eliminating the fluctuations of the unwanted
$\varphi$ and $\chi$ fields.
In other words, the modified version of the simplified BLG
action of \cite{M} is now \cite{triple}:
\be
-\frac{1}{2}{\rm tr}\Big(D_\mu \phi^I - B_\mu\varphi^I\Big)^2
+\frac{i}{2} {\rm tr} \psi^A\hat D \Big(\psi^A - \hat B\chi^A\Big)
+ \nn \\ +
\Big(\partial_\mu\tilde\varphi^I - {\rm tr}( B_\mu \phi^I)\Big)
\partial_\mu\varphi^I
 -\frac{i}{2}\bar{\tilde\chi}^A \hat\partial\chi^A -
\frac{i}{2}\bar\chi^A\Big(\hat \partial\tilde\chi^A -
{\rm tr}(B_\mu \psi^A)\Big) + \nn \\
+ \frac{1}{2}\epsilon^{\mu\nu\lambda} {\rm tr}
\Big(F_{\mu\nu}B_\lambda\Big) -
\frac{1}{12}{\rm tr} \Big(\varphi^I[\phi^J,\phi^K] +
\varphi^j[\phi^K,\phi^I] + \varphi^K[\phi^I,\phi^J]\Big)^2 +\nn\\
+ \frac{i}{2}\Gamma_{AB}^{IJ}\varphi^I
{\rm tr}\Big(\bar\psi^A[\phi^J,\psi^B]\Big)
+\frac{i}{4}\Gamma^{IJ}_{AB} {\rm tr}\Big(\bar\psi^A
[\phi^I,\phi^J]\Big)\chi^B
-\frac{i}{4}\Gamma_{AB}^{IJ} \bar\chi^A {\rm tr}
\Big([\phi^I,\phi^J]\psi^B\Big)
\label{siBLG}
\ee
It essentially differs from eq.(21) of \cite{M} in the second
line.\footnote{A trivial mistake of \cite{M}
in the $\phi^6$ term (omitted item $2\sum_{I<J}\varphi^I\varphi^J
{\rm tr}\Big([\phi^I,\phi^K][\phi_j,\phi^K]\Big)$
is also corrected in \cite{triple} and in (\ref{siBLG}).}
Here $\phi^I$ and $\psi^A$ with $I=1,\ldots,8$ and $A=1,\ldots,8$
are real- and Grassmann-valued elements of the vector and spinor
representations of the $SO(8)$ group respectively (related by
octonionic triality to the second spinor representation, where
the $N=8$ SUSY transformation parameter takes values).
They are also $N\times N$ matrices, i.e. belong to adjoint
representation of the gauge group $G=SU(N)$.
$A_\mu$ is the corresponding connection, also in the adjoint of $G$,
$D_\mu \phi = \partial_\mu\phi - [A_\mu,\phi]$, $F_{\mu\nu} =
\partial_\mu A_\nu - \partial_\nu A_\mu + [A_\mu,A_\nu]$,
and $B_\mu$ is an auxiliary adjoint vector field (not a connection).
The other auxiliary fields $\varphi^I,\tilde\varphi^I$ and their
superpartners $\chi^A,\tilde\chi^A$ are $G$-singlets (are color-less),
possibly fragments of Kac-Moody extension of $G$.
See \cite{M,triple} for further details.

\bigskip

As explained in \cite{triple},

$\bullet$ The action (\ref{siBLG})
is $N=8$ supersymmetric due to original BL theorem \cite{BL},
because it is now based on the 3-algebra with totally
antisymmetric structure constants, which is a central extension
of the degenerate one \cite{triNa} used in \cite{M}:
\be
\left[X,Y,Z\right] = {\rm tr}\, X \cdot [Y,Z] +
{\rm tr}\, Y \cdot [Z,X] + {\rm tr}\, Z \cdot [X,Y] +
\zeta \cdot {\rm tr}\left( X [Y,Z]\right)
\label{Nabr}
\ee
$\zeta$ is a central element, different from unity matrix $I$
and related to it by non-trivial scalar product
$<\!I,I\!>\, = \,<\!\zeta,\zeta\!>\, =0$,
$\ <\!I,\zeta\!>\, =\, <\!\zeta,I\!>\, = -1$,
so that the 3-algebra metric is
$\left(\begin{array}{cc|c}
0&-1&0\\-1&0&0\\ \hline  0&0& h
\end{array}\right)$ and 
$f^{abcI} = -{f^{abc}}_\zeta = -f^{abc} = -f^{Iabc}$.
The last term of (\ref{Nabr}) was absent in \cite{M}
and this made the 3-bracket degenerate and the structure constants
(with the forth index raised by 3-algebra metric)
not totally antisymmetric.
The $\varphi,\chi$ fields are associated with the $I$
(matrix-trace) generator, while $\tilde\varphi,\tilde\chi$
-- with the central element $\zeta$.
Non-trivial 3-algebra metric implies the $\varphi$-$\tilde\varphi$
and $\chi$-$\tilde\chi$ mixing form of the kinetic terms in
(\ref{siBLG}).

$\bullet$ The would-be coupling constant in front of
non-quadratic terms in (\ref{siBLG}) can be absorbed into
rescaling of $\varphi$ and $\chi$, accompanied by rescaling
of $\tilde\varphi$ and $\tilde\chi$ in the opposite direction.
This lack of this feature was one of the problems in \cite{M},
and in (\ref{siBLG}) we have an action, which has no dimensionless
coupling constants, as required in M-theory.

$\bullet$ All the unwanted extra fields
$\varphi,\chi,\tilde\varphi,\tilde\chi$ are auxiliary:
they do not propagate and contribute only through boundary
terms\footnote{In this respect the action in the
$\varphi$-$\tilde\varphi$ sector is reminiscent of the one,
considered in \cite{Kro}.}
(i.e. to correlators) and zero-modes.

$\bullet$ The fact that Lagrange multiplier $\tilde\varphi$
nullifies only $\partial^2\varphi$ rather than $\varphi$ itself
is very important, because this allows the zero-mode
$\varphi=const$.
Among other effects, this zero mode can form
a condensate, producing a term
$<\!\varphi\!>^2\!{\rm tr}\, B^2$ 
from the first item in (\ref{siBLG}),
which, after auxiliary field $B_\mu$ is integrated away,
converts the Chern-Simons interaction ${\rm tr} F\wedge B$
into kinetic Yang-Mills term
$<\!\varphi\!>^{-2}\! {\rm tr}\, F_{\mu\nu}^2$ for connection $A_\mu$.
This means that despite $\varphi$ fields are now auxiliary,
the crucially important possibility to use them for the
M2 $\rightarrow$ D2 conversion {\it a la} \cite{Mu} is preserved.

\bigskip

All this means that today we possess a perfectly simple version
(\ref{siBLG}) of the BLG action for arbitrary number of M2
branes, there are no longer doubts about its existence for
arbitrary gauge group $SU(N)$, there are no coupling constants,
no extra dynamical fields, and it is clearly related to the other
brane actions, as required by embeddings of $d=10$ superstring
models into the $d=11$ M-theory.
The road is now open for building up the first-quantized theory
of M2 branes (supermembranes).
This implies that attention can now be shifted from the study of
3-algebra structure (where a lot of interesting questions still
remain) to the other issues:
we know what should be the crucial next steps from the history
of first-quantized theory of superstrings.

Constructing the action (\ref{siBLG}) can be considered 
as the very first step,
corresponding to substitution of Nambu-Goto action for bosonic
strings by a $\sigma$-model action, of which (\ref{siBLG})
is supposed to be a (super)membrane analogue.
In the case of membranes
the problem was more complicated, because Nambu-Goto action is
ill (does not damp fluctuations) from the very beginning,
no approach to bosonic membrane is still available (problems look
more severe than the tachyon of bosonic string) and one should
begin directly from the supersymmetric case, moreover supersymmetry
should be immediately extended to ${\cal N}=8$.
Thus it may be not too surprising that we had to wait till
2008 to have this action written down...

In the case of strings the next big step was consideration of
world sheets with non-trivial topologies, with two complementary
formalisms finally developed for this purpose
(and still not fully related,
see \cite{LM} for description of the corresponding problems).
One is the Polyakov formalism \cite{Po},
promoting the $\sigma$-model action to arbitrary curved
$2d$ geometries and generalizing the treatment of relativistic
particle in \cite{BDH}.
Another is equilateral-triangulation approach, nicely expressed
in terms of matrix models \cite{mamos}
and formally equivalent to substitution
of smooth $2d$-geometries by Grothendieck's
{\it dessins d'enfants} \cite{Gro}.
In the case of membranes this step is going to be a hard exercise,
already because the topological classification of $3d$ world
volumes is far more complicated than in $2d$.
Still, the very first movement -- introduction of $3d$ geometry 
into (\ref{siBLG}) by both above-mentioned methods --
should be straightforward, and undoubtedly very interesting.
For a variety of reasons it seems natural to do this in the
modern BF-version of Palatini formalism, which is now widely
popularized by controversial, but inspiring papers of G.Lisi
\cite{Lisi}.
Of certain help can be also comparison with the Green-Schwarz
formalism for the superstrings \cite{GSA}, where world-sheet
action has some common features with (\ref{siBLG}): it also
looks non-linear, but actually non-linearities concern only the
zero-modes and boundary effects.

Of crucial importance should be identification of the relevant
world-sheet-geometry degrees of freedom (moduli), which
the action is going to depend upon. This is not the $3d$
metric or dreibein and spin-connection themselves -- already
because of the general covariance. However, as we know
from experience with strings, the remaining degrees of freedom
(Liouville field) can also be irrelevant (or identified with
the other physical fields \cite{DK}), so that the only remaining
moduli are those of the $2d$ complex structures: finitely many
for any given $2d$ topology.
It is the analogue of Belavin-Knizhnik theorem \cite{BK,BKoth}
that formulates this statement for strings, which should be the
next big discovery in the story of BLG actions.
Again, there are many complications in the case of membranes:
as already mentioned, from the very beginning we need
supersymmetry (and the corresponding problem for superstrings was
partly resolved only quite recently! \cite{Ie,DHP}). Moreover,
the analogue of Riemann theta-function theory \cite{rtheta}
in $3d$ is not yet at our disposal -- and here we should face
the same problems as the other approaches to $3d$ topological
theories \cite{Ok}: there are no conventional terms to express
our answers through...

In this short summary we do not speculate about the resolution
of all these problems, i.e. about filling the empty spaces
in the right column of the following table:

\bigskip

\centerline{
\begin{tabular}{|c|c|}
\hline & \\
(super)strings & (super)membranes \\
&\\
\hline\hline & \\
$2d$ Nambu-Goto action $\rightarrow$ $2d$ $\sigma$-model action &
spirit of membrane $\rightarrow$
simplified BLG action (\ref{siBLG})\\
&\\
\hline & \\
Polyakov formalism:   & BF-version of Palatini formalism in $3d$\\
introduction of $2d$ metric, &\\
critical dimensions (where massless excitations occur), &\\
sum over geometries, & \\
sum over topologies,   & \\
relation to equilateral triangulations approach & \\
&\\
\hline & \\
Belavin-Knizhnik theorem: & \\
reduction of sum over metrics to sum over moduli &\\
&\\
\hline & \\
topology of world sheet:  &\\
spin structures and GSO projection, &\\
string field theory, &\\
boundary correlators and AdS/CFT correspondence & \\
&\\
\hline & \\
topology of the target space (compactifications): &\\
generic $2d$ conformal theories, &\\ $T$-dualities, &\\
other dualities & \\
&\\
\hline & \\
\ldots & \\
&\\
\hline
\end{tabular}
}

\bigskip \bigskip

\noindent
Our goal is to emphasize that we are
now in front of the new and interesting breakthrough into
the unknown -- the possibility opened to us by the timely
formulated problem \cite{JS1}, a brilliant insight \cite{BL,G}
and qualified polishing \cite{BLGfirst}-\cite{BLGlast},
culminated in \cite{triple} in the elegant formula 
(\ref{siBLG}), which is going to be -- perhaps, in some
reshaped and redecorated version -- a new focus of attention
in string theory in the coming years.

\section*{Acknowledgements}

This work  is partly supported by Russian Federal Nuclear
Energy Agency and Russian Academy of Sciences,
by the joint grant 06-01-92059-CE,  by NWO project 047.011.2004.026,
by INTAS grant 05-1000008-7865, by ANR-05-BLAN-0029-01 project,
by RFBR grant 07-02-00645 and
by the Russian President's Grant of Support for the Scientific
Schools NSh-3035.2008.2

\end{document}